# Interferenceless Polarization Splitting through Nanoscale van der Waals Heterostructures

Shahnawaz Shah,[1,2] Xiao Lin,[3,*] Lian Shen,[1,2] Maturi Renuka,[1,2] Baile Zhang,[3,4,*] and Hongsheng Chen[1,2,*]

[1] *Key Laboratory of Advanced Micro/Nano Electronic Devices & Smart Systems of Zhejiang, Zhejiang University, Hangzhou 310027, China.*

[2] *The Electromagnetics Academy at Zhejiang University, College of Information Science and Electronic Engineering, Zhejiang University, Hangzhou 310027, China.*

[3] *Division of Physics and Applied Physics, School of Physical & Mathematical Sciences, Nanyang Technological University, Singapore 637371, Singapore.*

[4] *Centre for Disruptive Photonic Technologies, NTU, Singapore 637371, Singapore.*

[*] *Corresponding Authors: xiaolinbnwj@ntu.edu.sg (X. Lin); bailezhang@ntu.edu.sg (B. Zhang); hansomchen@zju.edu.cn (H. Chen).*

**ABSTRACT:** The ability to control the polarization of light at the extreme nanoscale has long been a major scientific and technological goal for photonics. Here we predict the phenomenon of polarization splitting through van der Waals heterostructures of nanoscale thickness, such as graphene-hexagonal boron nitride ($h$BN) heterostructures, at infrared frequencies. The underlying mechanism is that the designed heterostructures possess an effective relative permittivity with its in-plane (out-of-plane) component being unity (zero); such heterostructures are transparent to the transverse-electric (TE) waves while opaque to the transverse-magnetic (TM) waves, without resorting to the interference effect. Moreover, the predicted phenomenon is insensitive to incident angles. Our work thus indicates that van der Waals heterostructures are a promising nanoscale platform for the manipulation of light, such as the design of polarization beam nano-splitters and epsilon-near-zero materials, and the exploration of superscattering for TM waves while zero scattering for TE waves from deep-subwavelength nanostructures.

## I. INTRODUCTION

The emergence of layered two-dimensional materials [1-3] and van der Waals heterostructures [4, 5], such as graphene [6-11], hexagonal boron nitride (*h*BN) [12-19] and graphene-*h*BN heterostructures [20-25], has ignited numerous nanophotonic studies of both fundamental and applied nature. Due to their appealing optical properties, people have tried to control the flow of light at the nanoscale and even at the atomic scale [1, 2, 6, 7, 26]. For example, it was recently reported possible to design the surface cloaks [27] and perfect absorbers [28] using graphene monolayer, and to realize the negative refraction in graphene-*h*BN heterostructures with a thickness of several nanometers [24]. One remaining open challenge is to control the polarization of light at the extreme nanoscale, especially to achieve the total transmittance for one linearly polarized wave (such as TE or *s*-polarized wave) while the total reflectance for the other linearly polarized wave (TM or *p*-polarized wave) within a wide range of incident angles. Below, this exotic phenomenon is referred as the perfect polarization splitting, which is highly sought after but has not been realized through nanostructures without external magnetic fields [26]. The perfect polarization splitting is of paramount importance to photonics. This is because that polarization is an inherent characteristic of light, and its flexible control can result in many intriguing phenomena and unique applications [29, 30]. One typical application of polarization splitting is the polarization beam splitters, which have been widely used in photonic systems, including optical communications, imaging processing and integrated photonic circuits [30-33]. Conventional polarization beam splitters are designed based on the birefringence of anisotropic materials [33], Brewster effect [33, 34], light interference in optically thin films [33], photonic crystals [35, 36], metamaterials [37, 38], metasurfaces [39], and so on. Although these polarization beam splitters can have good performance in the separate control of differently polarized light, they still require a thickness in the order of wavelength scale (e.g., around one-tenth of the wavelength based on metasurfaces [39]).

In this work, we predict the polarization splitting of linearly polarized waves through van der Waals heterostructures, such as graphene-*h*BN heterostructures, with a nanoscale thickness (<10 nm for the example given below) at the infrared regime (near 25.35 THz). In other words, for the ideal lossless case, the total reflectance of TM waves and the total transmittance of TE waves can be simultaneously achieved, for arbitrary incident angle. This is because the graphene-$h$BN heterostructures can be designed to have an effective permittivity with its in-plane (out-of-plane) component being unity (zero). This way, the designed heterostructures is transparent to TE waves while opaque to TM waves, without the help of interference effect. We note that the interferenceless effect was once used to achieve the perfect absorption by a van der Waals crystal [40, 41]; our work shall be a further development of the interferenceless effect [40, 41] in the control of polarization of light. With the consideration of realistic material loss, the polarization splitting, i.e., with a high ratio (>100) of reflectance or transmittance between different polarized light, can still be maintained within a wide range of incident angles near 25.35 THz.

## II. Perfect polarization splitting of TM and TE waves

We start with the schematic illustration of the phenomenon of perfect polarization splitting for TM and TE waves in Fig. 1. The wave is incident from region 1 with a relative permittivity of $\varepsilon_{1,r} = 1$ (i.e., air) and transmits into region 2 with $\bar{\bar{\varepsilon}}_{2,r} = [\varepsilon_{\parallel,r}, \varepsilon_{\parallel,r}, \varepsilon_{\perp,r}]$. From the classic electromagnetic theory [42], the reflection coefficients for TM and TE waves are derived as

$$R_{1|2}^{TM} = \frac{\cos\theta\, \varepsilon_{\parallel,r} - \sqrt{\varepsilon_{\parallel,r} - \frac{\varepsilon_{\parallel,r}}{\varepsilon_{\perp,r}}\sin^2\theta}}{\cos\theta\, \varepsilon_{\parallel,r} + \sqrt{\varepsilon_{\parallel,r} - \frac{\varepsilon_{\parallel,r}}{\varepsilon_{\perp,r}}\sin^2\theta}} \tag{1}$$

$$R_{1|2}^{TE} = \frac{\cos\theta - \sqrt{\varepsilon_{\parallel,r} - \sin^2\theta}}{\cos\theta + \sqrt{\varepsilon_{\parallel,r} - \sin^2\theta}} \tag{2}$$

where $\theta$ is the incident angle. Correspondingly, the transmission coefficients for TM and TE waves are $T_{1|2}^{TM} = \frac{2\varepsilon_{\parallel,r}\cos\theta}{\cos\theta\, \varepsilon_{\parallel,r} + \sqrt{\varepsilon_{\parallel,r} - \frac{\varepsilon_{\parallel,r}}{\varepsilon_{\perp,r}}\sin^2\theta}}$ and $T_{1|2}^{TE} = \frac{2\cos\theta}{\cos\theta + \sqrt{\varepsilon_{\parallel,r} - \sin^2\theta}}$, respectively; for the ideal lossless case, the transmittance of TE waves is equal to $t^{TE} = 1 - |R_{1|2}^{TE}|^2$. When region 2 is an isotropic medium (i.e.,

$\varepsilon_{\parallel,r} = \varepsilon_{\perp,r}$), equations (1-2) indicate that there will be nonzero reflectance for both TM and TE waves (see Fig. 1(a)), except when the incident angle is the Brewster angle for TM waves [34, 42].

When region 2 is a uniaxial medium (i.e., $\varepsilon_{\parallel,r} \neq \varepsilon_{\perp,r}$), we find that it is possible to achieve the perfect polarization splitting with $|R_{1|2}^{TM}| = 1$ and $R_{1|2}^{TE} = 0$, i.e., total reflectance (transmittance) for TM (TE) waves, for arbitrary incident angle; see Fig. 1(b). From equations (1-2), the corresponding condition is $\varepsilon_{\parallel,r} = 1$ and $\varepsilon_{\perp,r} = 0$. To our knowledge, this condition has not been directly discussed for the realization of perfect polarization splitting. We note that Ref. [37] also used uniaxial metamaterials to obtain the perfect polarization splitting. However, while the optical axis of our uniaxial medium is perpendicular to the interface (i.e., along z direction), the optical axis of their uniaxial medium [37] is parallel to the interface (i.e. along y direction). Consequently, their perfect polarization splitting can be achieved only at certain values of $\phi$ (i.e., $\phi = 0$ in Ref. [37]), where $\phi$ is the angle between the incident plane (which contains the surface normal, i.e., the z axis, and the propagation wavevector) and the plane containing the surface normal and the optical axis of uniaxial medium(i.e., the y-z plane). In contrast, our proposed structure has no such limit, i.e., the perfect polarization splitting is independent of $\phi$, due to the rotational symmetry of our structure with respect to the surface normal. In addition, our work is different from Refs. [40, 41], which discussed the condition to achieve $|R_{1|2}^{TE}| = 0$ or $|R_{1|2}^{TM}| = 1$, i.e., total reflectance for TM waves or total transmittance for TE waves at *different* frequencies, instead of the same frequency.

Figure 2 numerically verifies the perfect polarization splitting of TM and TE waves by showing the reflectance. At the interface of air/semi-infinite uniaxial medium with $\bar{\bar{\varepsilon}}_{2,r} = [1, 1, 0]$, Fig. 2(a) shows $\left|R_{1|2}^{TE}\right|^2 = 0$ and $\left|R_{1|2}^{TM}\right|^2 = 1$ for arbitrary value of incident angles. Moreover, the perfect polarization splitting can also be achieved through a thin slab of such uniaxial medium; see Fig. 2(b). For conceptual illustration, we assume the uniaxial slab with a thickness of $d$ is surrounded by air, i.e., $\varepsilon_{3,r} = \varepsilon_{1,r} = 1$.

This way, the reflection and transmission coefficients of TE and TM waves through a uniaxial slab are derived as [42]

$$R_{1|3}^{TM} = R_{1|2}^{TM} + \frac{T_{1|2}^{TM} R_{2|3}^{TM} T_{2|1}^{TM} e^{2ik_{z,2}^{TM} d}}{1 - R_{2|1}^{TM} R_{2|3}^{TM} e^{2ik_{z,2}^{TM} d}} \quad (3)$$

$$T_{1|3}^{TM} = \frac{T_{1|2}^{TM} T_{2|3}^{TM} e^{ik_{z,2}^{TM} d}}{1 - R_{2|1}^{TM} R_{2|3}^{TM} e^{2ik_{z,2}^{TM} d}} \quad (4)$$

$$R_{1|3}^{TE} = R_{1|2}^{TE} + \frac{T_{1|2}^{TE} R_{2|3}^{TE} T_{2|1}^{TE} e^{2ik_{z,2}^{TE} d}}{1 - R_{2|1}^{TE} R_{2|3}^{TE} e^{2ik_{z,2}^{TE} d}} \quad (5)$$

$$T_{1|3}^{TE} = \frac{T_{1|2}^{TE} T_{2|3}^{TE} e^{ik_{z,2}^{TE} d}}{1 - R_{2|1}^{TE} R_{2|3}^{TE} e^{2ik_{z,2}^{TE} d}} \quad (6)$$

In above equations, $T_{2|1}^{TM} = T_{2|3}^{TM} = \frac{2\sqrt{\varepsilon_{\parallel,r} - \frac{\varepsilon_{\parallel,r}}{\varepsilon_{\perp,r}} \sin^2\theta}}{\cos\theta\, \varepsilon_{\parallel,r} + \sqrt{\varepsilon_{\parallel,r} - \frac{\varepsilon_{\parallel,r}}{\varepsilon_{\perp,r}} \sin^2\theta}}$; $R_{2|1}^{TM} = R_{2|3}^{TM} = -R_{1|2}^{TM}$; $T_{2|1}^{TE} = T_{2|3}^{TE} = \frac{2\sqrt{\varepsilon_{\parallel,r} - \sin^2\theta}}{\cos\theta + \sqrt{\varepsilon_{\parallel,r} - \sin^2\theta}}$; $R_{2|1}^{TE} = R_{2|3}^{TE} = -R_{1|2}^{TE}$; $k_{z,2}^{TM} = \frac{\omega}{c}\sqrt{\varepsilon_{\parallel,r} - \frac{\varepsilon_{\parallel,r}}{\varepsilon_{\perp,r}}\sin^2\theta}$; $k_{z,2}^{TE} = \frac{\omega}{c}\sqrt{\varepsilon_{\parallel,r} - \sin^2\theta}$; $\omega$ is the angular frequency; and $c$ is the light speed in free space.

Figure 2(b) shows the reflectance of TM and TE waves through a uniaxial slab with $\bar{\bar{\varepsilon}}_{2,r} = [1, 1, 0]$. For an arbitrary incident angle (such as $30^\circ$ in Fig. 2(b)), the reflectance of TM waves is unity (i.e., $r^{TM} = |R_{1|3}^{TM}|^2 = 1$) and the reflectance of TE waves is zero (i.e., $r^{TE} = |R_{1|3}^{TE}|^2 = 0$); both $r^{TM}$ and $r^{TE}$ are independent of the slab thickness. Such behavior can be explained by the interferenceless effect. For $R_{1|3}^{TM}$ in equation (3), since $T_{1|2}^{TM} = 0$ for arbitrary incident angle, one has $r^{TM} = |R_{1|3}^{TM}|^2 = |R_{1|2}^{TM}|^2 = 1$, which is independent of the slab thickness $d$. Similarly, for $R_{1|3}^{TE}$ in equation (5), since $R_{1|2}^{TE} = -R_{2|3}^{TE} = 0$ for arbitrary incident angle, one has $r^{TE} = |R_{1|3}^{TE}| = 0$ (and thus $t^{TE} = 1 - r^{TE} = 1$), which is also independent of the slab thickness $d$. This underlying new mechanism without resorting to the interference effect, different from previous works [33-39], may provide the possibility to design the ultrathin polarization splitter of nanoscale thickness; see below discussions.

## III. RESULTS AND DISCUSSION

Figure 3 shows graphene-$h$BN heterostructures are a viable platform to construct the uniaxial medium with $\bar{\bar{\varepsilon}}_{2,r} = [1, 1, 0]$ near $h$BN's first reststrahlen band (i.e., at 25.35 THz). Figure 3(a) schematically shows the heterostructure, where the unit cell contains a monolayer graphene and a $h$BN slab with a thickness of $d_{hBN} = 9$ nm. The experimental data of isotopically enriched $h$BN slab is adopted to model $h$BN's relative permittivity [43, 44], i.e., $\bar{\bar{\varepsilon}}_{r,hBN} = [\varepsilon_{\parallel,hBN}, \varepsilon_{\parallel,hBN}, \varepsilon_{\perp,hBN}]$; see appendix. Since the studied wave has the component of wavevector parallel to the interface less than $\omega/c$, the nonlocal effect of graphene is minor and can be neglected [45]; the Kubo formula is thus adopted to model the surface conductivity $\sigma_s$ of graphene [45], see appendix; graphene has a chemical potential of $\mu_c = 0.1484$ eV, and a conservative electron mobility of 30000 cm$^2$V$^{-1}$s$^{-1}$ [46, 47], which characterizes the loss in graphene, is assumed. This way, the effective relative permittivity of graphene can be described by $\bar{\bar{\varepsilon}}_{r,gra} = [\varepsilon_{\parallel,gra}, \varepsilon_{\parallel,gra}, \varepsilon_{\perp,gra}]$, where $\varepsilon_{\parallel,gra} = 1 + \frac{i\sigma_s}{\omega\varepsilon_0 d_{gra}}$ [6] and $\varepsilon_{\perp,gra} = 3$ [48], the thickness of graphene is $d_{gra} = 0.35$ nm [49], and $\varepsilon_0$ is the permittivity of free space.

Since the thickness of the unit cell of graphene-$h$BN heterostructure, i.e., $d_0 = d_{gra} + d_{hBN} = 9.35$ nm, is much smaller than the wavelength of light considered ($\lambda > 10$ μm), the effective medium theory [42] is applicable to model the effective relative permittivity $\bar{\bar{\varepsilon}}_{r,eff} = [\varepsilon_{\parallel,eff}, \varepsilon_{\parallel,eff}, \varepsilon_{\perp,eff}]$ of the designed heterostructures, where

$$\varepsilon_{\perp,eff} = \frac{\varepsilon_{\perp,gra}\varepsilon_{\perp,hBN}(d_{gra}+d_{hBN})}{d_{gra}\varepsilon_{\perp,hBN}+d_{hBN}\varepsilon_{\perp,gra}} \tag{7}$$

$$\varepsilon_{\parallel,eff} = \frac{\varepsilon_{\parallel,gra} \cdot d_{gra} + \varepsilon_{\parallel,hBN} \cdot d_{hBN}}{d_{gra} + d_{hBN}} \tag{8}$$

Since $\text{Re}(\varepsilon_{\perp,hBN}) = 0$ at 25.35 THz, equation (7) directly indicates $\text{Re}(\varepsilon_{\perp,eff}) \approx 0$ at 25.35 THz; this also indicates that the value of $\varepsilon_{\perp,gra}$ has minor influence in the performance of our designed polarization splitting. Since $\varepsilon_{\parallel,gra} < 0$ and $\varepsilon_{\parallel,hBN} > 0$ at 25.35 THz, it is feasible to achieve $\text{Re}(\varepsilon_{\parallel,eff}) = 1$ at 25.35 THz via tailoring the chemical potential of graphene from equation (8). By following the above design procedure, Fig. 3(b) shows that $\text{Re}(\varepsilon_{\perp,\text{eff}}) = 0$ and $\text{Re}(\varepsilon_{||,\text{eff}}) = 1$ are simultaneously achieved at 25.35 TH; due to the existence of material loss, one also has $\text{Im}(\varepsilon_{\perp,\text{eff}}) = 0.0225$ and $\text{Im}(\varepsilon_{||,\text{eff}}) = 0.2462$ at 25.35 THz.

Figure 4 shows the performance of polarization splitting of TM and TE waves at the interface of air/semi-infinite graphene-$h$BN heterostructures. For the simplicity of conceptual demonstration, we first neglect the material loss, i.e., by artificially neglecting the imaginary part of permittivity. Figure 3(a-b) shows the reflectance of TM and TE waves, respectively; Figure 3(c) shows the ratio of reflectance between TM and TE waves. As expected, the perfect polarization splitting is achieved at 25.35 THz, independent of the incident angle. Moreover, when considering the realistic material loss, the polarization splitting with good performance ($|R_{1|3}^{\text{TM}}|^2/|R_{1|3}^{\text{TE}}|^2 > 100$) is still maintained in a relative wide range of incident angles (i.e., [20° 30°]) near 25.35 THz; see Fig. 3(d). For the realistic lossy case, since all transmitted waves will be dissipated inside the semi-infinite heterostructures, only the reflectance ratio is suitable to describe the polarization splitting in Fig. 4. This also indicates our designed structures are suitable for the perfect absorption of TM waves [40, 41].

Figures 5-6 show the polarization splitting of TM and TE waves from an ultrathin slab of graphene-$h$BN heterostructures. In Figs. 5-6, the realistic material loss is considered. For the reflected waves in Fig. 5, the performance of polarization splitting can be improved by using a thinner slab of graphene-$h$BN heterostructure. Most importantly, the polarization splitting with $|R_{1|3}^{\text{TM}}|^2/|R_{1|3}^{\text{TE}}|^2 > 100$ can be achieved in a wide range of incident angles (i.e., [12° 90°]) near 25.35 THz, by only using an ultrathin slab with its thickness $d < 10$ nm (less than 1/1000 of the wavelength in free space, i.e., $d < \lambda/1000$); see Fig. 5(a). It is actually quite unexpected to see that the performance of polarization splitting from an ultrathin slab in Fig. 5(a) is much better than from a single interface in Fig. 4(d). For the transmitted waves in Fig. 6, the performance of polarization splitting is highly dependent on the slab thickness, and in contrast to Fig. 5, can be improved by using a thicker slab. In order to gain a better

performance, the thickness of the designed slab is required to be at least several hundred nanometers, such as $d \approx 250$ nm (still $d < \lambda/40$) in Fig. 6(c).

The above dependence of the performance of polarization splitting on the thickness of slab of graphene-$h$BN heterostructures in Figs. 5-6 is mainly due to the existence of materials loss. When the material loss exists, the value of $|R_{1|2}^{\text{TM}}| = |R_{2|3}^{\text{TM}}|$ will become non-unity, leading to $|R_{1|3}^{\text{TM}}| \neq 1$ in equation (3) and $|T_{1|3}^{\text{TM}}| \neq 0$ in equation (4); similarly, the value of $|R_{1|2}^{\text{TE}}| = |R_{2|3}^{\text{TE}}|$ will becomes nonzero, leading to $|R_{1|3}^{\text{TE}}| \neq 0$ in equation (5) and $|T_{1|3}^{\text{TE}}| \neq 1$ in equation (6). In other words, due to the material loss, there will be part of the incident field firstly transmitting into, then reflecting back and forth inside the loss uniaxial slab. This part of field will unavoidably degrade the performance of polarization splitting, especially for the transmitted waves.

## IV. CONCLUSION

In conclusion, we predict the interferenceless polarization splitting of linearly polarized waves with the use of van der Waals heterostructures. The underlying mechanism is that the heterostructures can be tailored to possess a relative permittivity with its in-plane (out-of-plane) component being unity (zero); then the heterostructures are intrinsically transparent (opaque) to TE (TM) waves. Most importantly, since the interferenceless mechanism does not rely on the interference effect, it provides a feasible way to achieve the polarization splitting via an ultrathin slab with its thickness down to even nanoscale. Therefore, our work indicates that the van der Waals heterostructures are a promising and versatile platform to manipulate light in the extreme nanoscale, such as the design of polarization nano-splitter and the widely-studied epsilon-near zero materials [50,51], and to demonstrate many other exotic optical phenomena. For example, it is highly wanted to simultaneously realize zero-scattering for TE waves and superscattering for TM waves [19] at the same frequency; this way, the detection will be highly sensitive to the polarization of light. This might be realized by a cylindrical subwavelength structure based on graphene-$h$BN heterostructure, which works near the frequency where the in-plane (out-of-plane) component of the effective relative permittivity of heterostructures is unity (negative [19]).

## APPENDIX: SURFACE CONDUCTIVITY OF GRAPHENE AND PERMITTIVITY OF *h*BN

In this work, Kubo formula [45] is adopted to model the surface conductivity of graphene monolayer, i.e., $\sigma_{\text{gra}} = \sigma_{\text{inter}} + \sigma_{\text{intra}}$, where $\sigma_{\text{inter}} = \frac{ie^2 k_B T}{\pi \hbar^2 \left(\omega + \frac{i}{\tau}\right)} \left[\frac{\mu_c}{k_B T} + 2\ln\left(e^{-\frac{\mu_c}{k_B T}} + 1\right)\right]$ and $\sigma_{\text{intra}} = \frac{ie^2(\omega + i/\tau)}{\pi \hbar^2} \int_0^\infty \frac{f_d(-x) - f_d(x)}{(\omega + i/\tau)^2 - 4\left(\frac{x}{\hbar}\right)^2} dx$. In above equations, $f_d(x) = \left(e^{\frac{(x-\mu_c)}{k_B T}} + 1\right)^{-1}$ is the Fermi-Dirac Distribution, $T = 300$ K, $e$ is the electron charge, $k_B$ is the Boltzman constant, and $\hbar$ is the reduced Plank's constant. The relaxation time is $\tau = \mu_c \mu/(ev_F^2)$, where the Fermi velocity is $v_F = 1 \times 10^6$ m/s. In this work, the chemical potential is set to be $\mu_c = 0.1484$ eV, and a conservative electron mobility of 30000 cm$^2$V$^{-1}$s$^{-1}$ [46, 47] is assumed; this way, the corresponding relaxation time is $\tau = 0.4452$ ps. The in-plane component of the relative permittivity of graphene monolayer is obtained via $\varepsilon_{\parallel,\text{gra}} = 1 + \frac{i\sigma_s}{\omega \varepsilon_0 d_{\text{gra}}}$ [6], where the thickness of monolayer graphene is $d_{\text{gra}} = 0.35$ nm [49]. The value of Re($\varepsilon_{\parallel,\text{gra}}$) is negative at 25.35 THz.

The uniaxial material *h*BN is characterized by a relative permittivity of $\bar{\bar{\varepsilon}}_{r,hBN} = [\varepsilon_{\parallel,hBN}, \varepsilon_{\parallel,hBN}, \varepsilon_{\perp,hBN}]$. The experimental data of isotopically enriched *h*BN slab [43, 44] is adopted to model *h*BN's relative permittivity, i.e., $\varepsilon_{l,hBN}(\omega) = \varepsilon_{l,\infty}\left(\frac{\omega_{l,\text{LO}}^2 - \omega^2 - i\Gamma_l \omega}{\omega_{l,\text{TO}}^2 - \omega^2 - i\Gamma_l \omega}\right)$, $l = \parallel, \perp$. For the in-plane direction ($l = \parallel$), the high-frequency permittivity $\varepsilon_{l,\infty} = 5.1$, the longitudinal optical frequency $\omega_{l,\text{LO}} = 1650$ cm$^{-1}$, the transverse optical frequency $\omega_{l,\text{TO}} = 1394.5$ cm$^{-1}$, and the phonon damping rate $\Gamma_l = 1.8$ cm$^{-1}$ [44]. For the out-of-plane direction ($l = \perp$), $\varepsilon_{l,\infty} = 2.5$, $\omega_{l,\text{LO}} = 845$ cm$^{-1}$, $\omega_{l,\text{TO}} = 785$ cm$^{-1}$, and $\Gamma_l = 1$ cm$^{-1}$ [44]. At 25.35 THz, one has $\varepsilon_{\perp,hBN} = 0$ and $\varepsilon_{\parallel,hBN} > 0$.

## ACKNOWLEDGMENTS
This work was sponsored by the National Natural Science Foundation of China under Grants No. 61625502, No. 61574127, No. 61601408, No. 61775193 and No. 11704332, the ZJNSF under Grant No. LY17F010008, the Top-Notch Young Talents Program of China, the Fundamental Research Funds for the Central Universities, and the Innovation Joint Research Center for Cyber-Physical-Society System, the Nanyang Technological University for Nanyang Assistant Professorship Start-Up Grant, the Singapore Ministry of Education under Grants No. MOE2015-T2-1-070 and No. MOE2011-T3-1-005, and Tier 1 RG174/16(S).

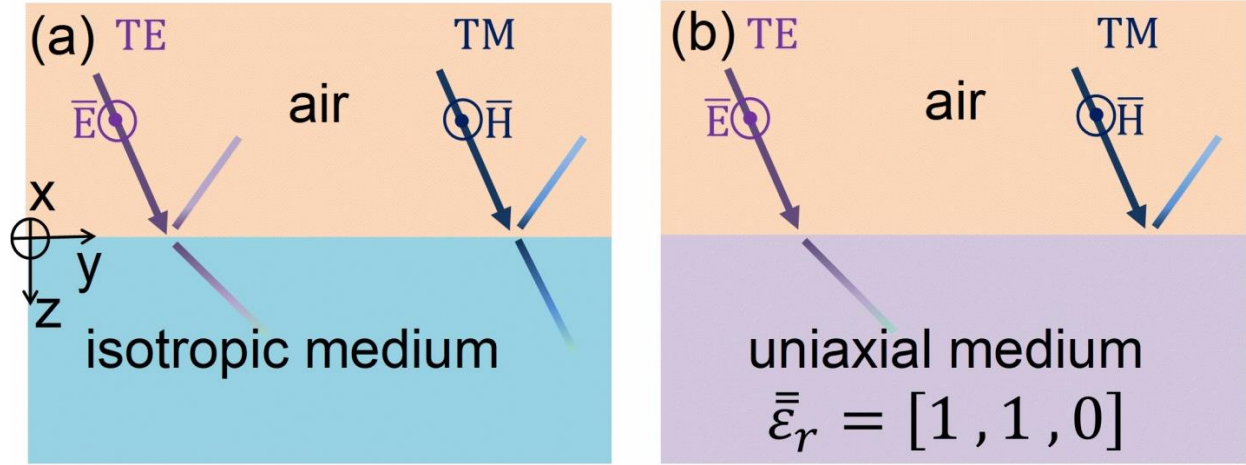

FIG. 1. Interferenceless perfect polarization splitting of TM and TE waves at a single interface. The reflectance and transmittance of TM and TE waves at the interface (a) of air/semi-infinite isotropic medium and (b) of air/semi-infinite uniaxial medium are schematically shown for comparison. The perfect polarization splitting, independent of the incidence angle, can be achieved via a uniaxial medium with a relative permittivity of $\bar{\bar{\varepsilon}}_r = [\varepsilon_{\parallel,r}, \varepsilon_{\parallel,r}, \varepsilon_{\perp,r}]$, where $\varepsilon_{\parallel,r} = 1$ and $\varepsilon_{\perp,r} = 0$.

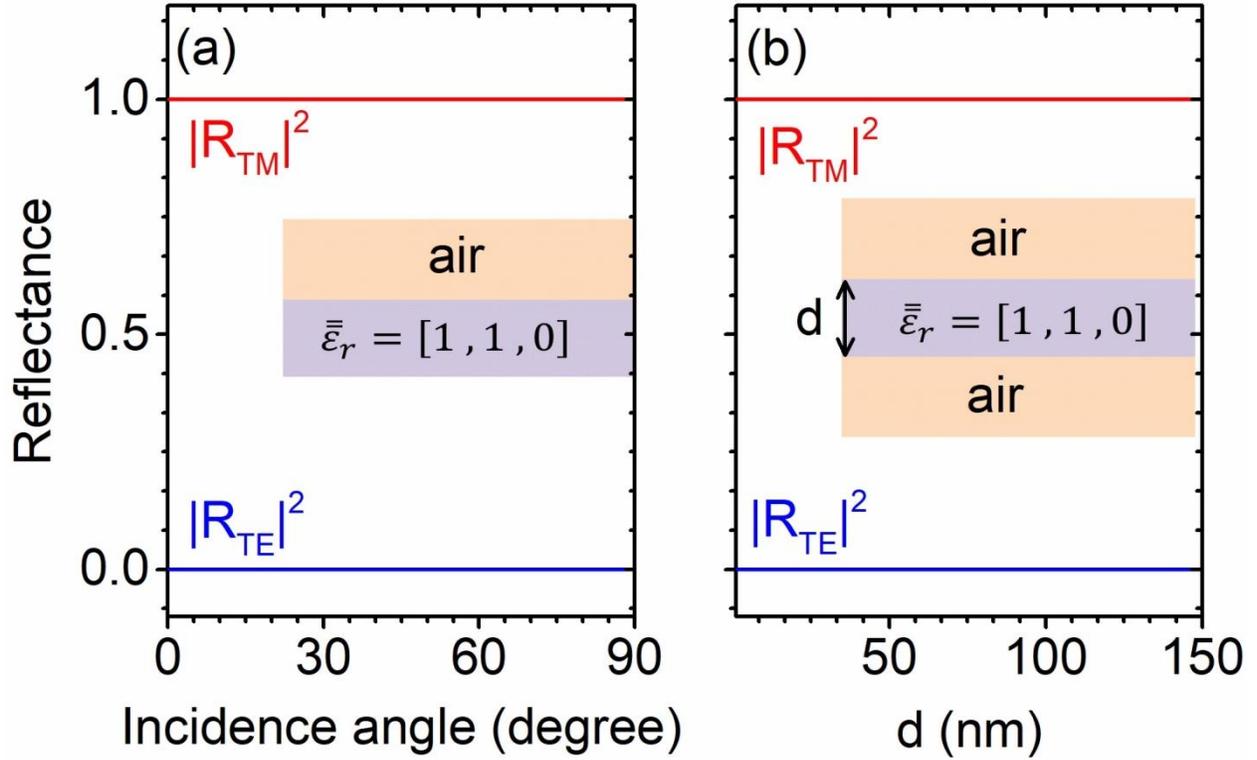

FIG. 2. Interferenceless perfect polarization splitting of TM and TE waves (a) at the interface of air/semi-finite uniaxial medium and (b) through a uniaxial slab with a thickness of $d$. For the perfect polarization splitting, the reflectance of TM waves is unity and the reflectance of TE waves is zero. The perfect polarization splitting is insensitive to the incidence angle and the slab thickness. The uniaxial medium has a relative permittivity of $\bar{\bar{\varepsilon}}_r = [1, 1, 0]$. The surrounding dielectric is air.

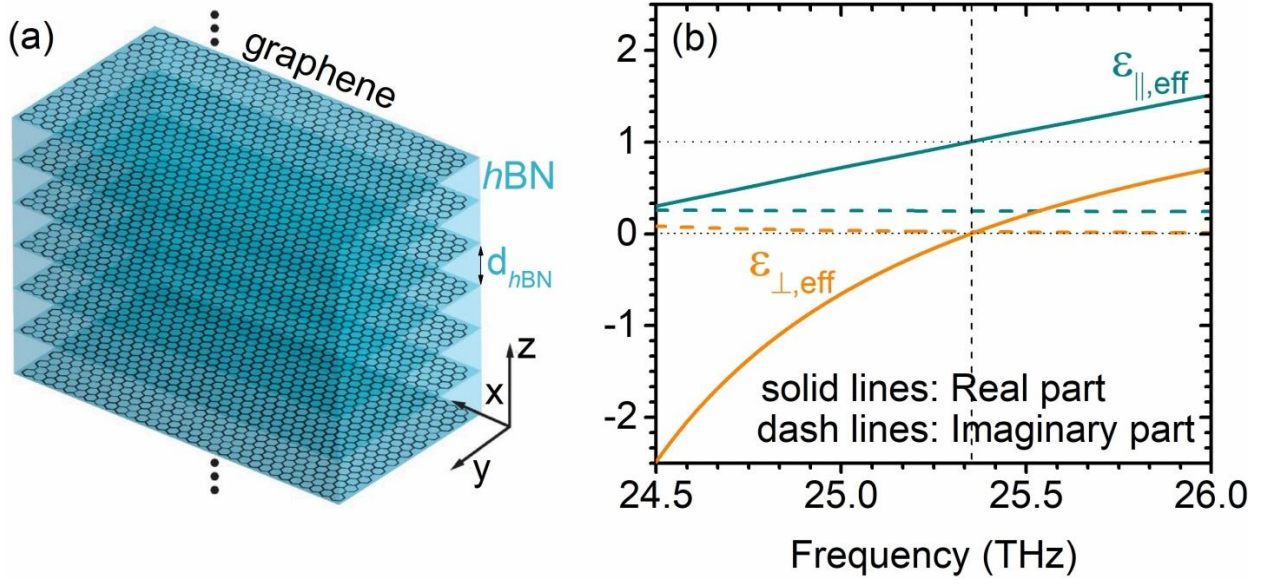

FIG. 3. Construction of the uniaxial medium with an effective relative permittivity of $[1, 1, 0]$ via graphene-$h$BN heterostructures. (a) Schematic illustration of graphene-$h$BN heterostructures. (b) Effective relative permittivity of graphene-$h$BN heterostructures, i.e., $\bar{\bar{\varepsilon}}_{\text{eff},r} = [\varepsilon_{\|,\text{eff}}, \varepsilon_{\|,\text{eff}}, \varepsilon_{\perp,\text{eff}}]$. The effective relative permittivity of $[1, 1, 0]$ is achieved at 25.35 THz. Here $d_{h\text{BN}} = 9$ nm; the monolayer graphene, with a thickness of 0.35 nm, has a chemical potential of $\mu_c = 0.1484$ eV and an electron mobility of 30000 cm$^2$V$^{-1}$s$^{-1}$.

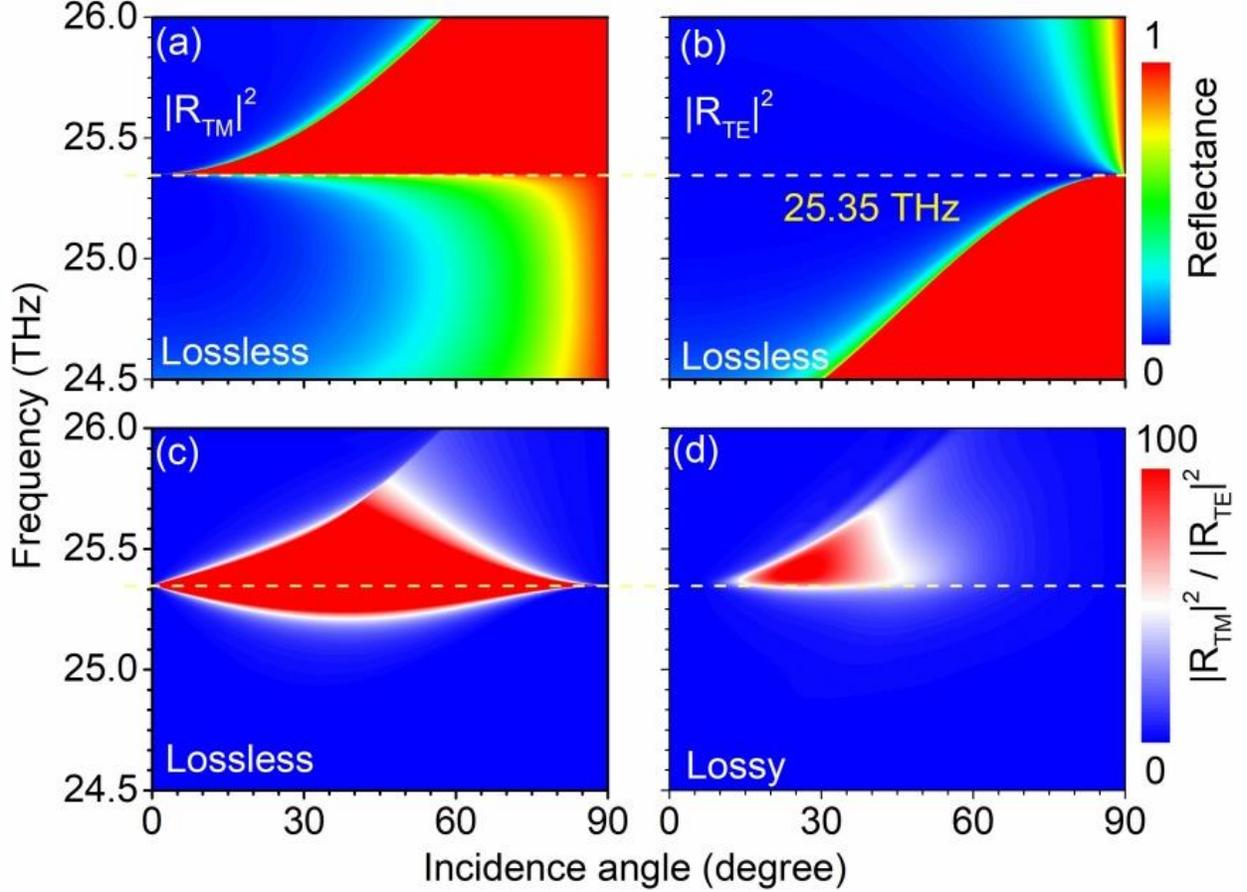

FIG. 4. Polarization splitting of TM and TE waves at the interface of air/semi-infinite graphene-$h$BN heterostructure. The setup of the heterostructure is the same as Fig. 3(b). Reflectance for (a) TM and (b) TE waves. (c) Ratio of reflectance between TM and TE waves. The perfect polarization splitting occurs at 25.35 THz. The material loss in (a-c) is neglected for conceptual illustration, i.e., the imaginary part of permittivity is artificially set to be zero. (d) The high ratio (>100) of reflectance between TM and TE waves is achieved in a relatively wide range of incidence angles (i.e., [20° 30°]) near 25.35 THz, when the realistic material loss is considered. For the realistic lossy case, all transmitted waves will be absorbed in the semi-infinite graphene-$h$BN heterostructure; this way, only the reflectance ratio is suitable to characterize the polarization splitting.

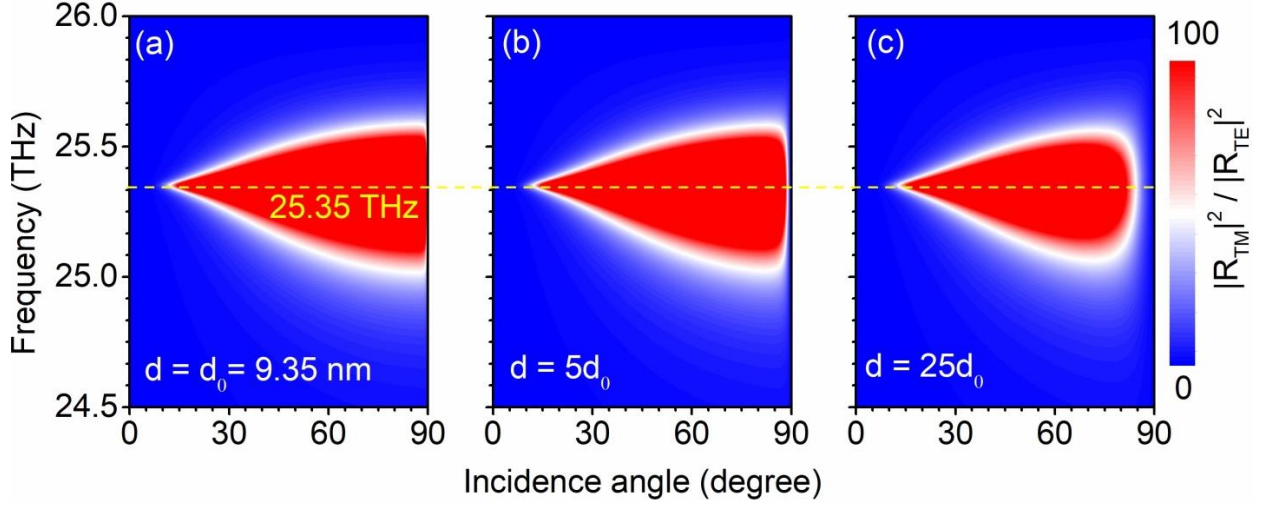

FIG. 5. Polarization splitting of TM and TE waves via a thin slab of graphene-hBN heterostructure. The graphene-hBN heterostructure, with its setup same as Fig. 3(b), has a thickness of $d$; the material loss is considered. Ratio of reflectance between TM and TE waves when $d$ is equal to (a) $d_0$, (b) $5d_0$, (c) $25d_0$, where $d_0 = 9.35$ nm. For the reflected waves, the performance of polarization splitting near 25.35 THz becomes better with the use of a thinner slab.

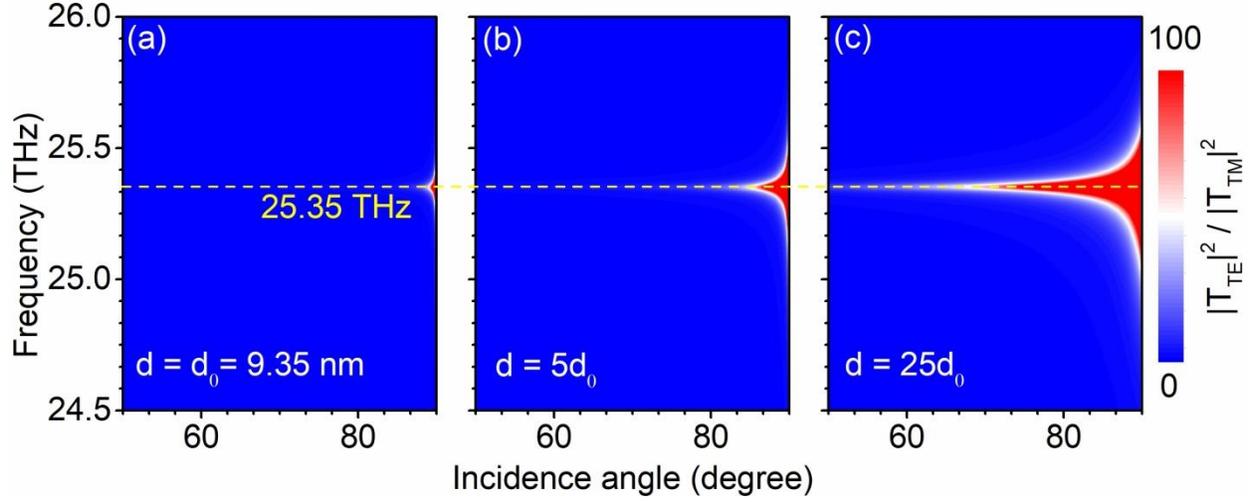

FIG. 6. Polarization splitting of TM and TE waves via a thin slab of graphene-$h$BN heterostructure. The setup of graphene-$h$BN heterostructures is the same as Fig. 5. Ratio of transmittance between TE and TM waves when $d$ is equal to (a) $d_0$, (b) $5d_0$, (c) $25d_0$. For the transmitted waves, the performance of polarization splitting near 25.35 THz becomes better with the use of a thicker slab.